\documentclass[sigconf,10pt]{acmart}
\renewcommand\footnotetextcopyrightpermission[1]{} 
\usepackage{booktabs} 

\theoremstyle{definition}

\newtheorem{example}{Example}

\usepackage{gensymb}

\usepackage[light]{Chivo}
\usepackage[T1]{fontenc}
\usepackage{xspace}
\usepackage{xcolor}
\usepackage{array,graphicx,subfig}

     \newcommand*{\origrightarrow}{}
     
     \renewcommand*{\textrightarrow}{\fontfamily{cmr}\selectfont\origrightarrow}

 \setcopyright{rightsretained}

 \acmDOI{10.475/123_4}

 \acmISBN{123-4567-24-567/08/06}

 \acmConference[WOODSTOCK'97]{ACM Woodstock conference}{July 1997}{El
   Paso, Texas USA}
 \acmYear{1997}
 \copyrightyear{2016}

 \acmArticle{4}
 \acmPrice{15.00}

 \editor{Jennifer B. Sartor}
 \editor{Theo D'Hondt}
 \editor{Wolfgang De Meuter}

\begin{document}
\setcopyright{none}
\title{MithraDetective:\\ A System for Cherry-picked Trendlines Detection}

\author{Yoko Nagafuchi$^*$, 
Yin Lin$^\dag$,
Kaushal Mamgain$^\P$,
Abolfazl Asudeh$^{**}$,\\
H. V. Jagadish$^\S$,
You (Will) Wu$^\parallel$,
Cong Yu$^\ddag$
}

\affiliation{%
 \institution{
 	$^{*,\dag,\S}$University of Michigan;
	$^{\P, **}$University of Illinois at Chicago; $^{\parallel, \ddag}$Google Research;
	\\
	\{yokon,irenelin,jag\}@umich.edu; \{kmamga2,asudeh\}@uic.edu;
	\{wuyou,congyu\}@google.com;
    }
\vspace{3mm}
}

\begin{abstract}
Given a data set, misleading conclusions can be drawn from it by cherry picking selected samples. One important class of conclusions is a trend derived from a data set of values over time.  Our goal is to evaluate whether the `trends' described by the extracted samples are representative of the true situation represented in the data. We demonstrate MithraDetective, a system to compute a support score to indicate how cherry-picked a statement is; that is, whether the reported trend is well-supported by the data.  The system
can also be used to discover more supported alternatives. MithraDetective provides an interactive visual interface for both tasks.
\end{abstract}

\keywords{Data Analysis, Data Fairness, Fact Checking, Computational Journalism}




\maketitle
\section{Introduction}
 Fake news has attracted much attention recently. While some fake news is just plain false, quite often we see impressive statements made on the basis of cherry-picked data points, with a particular agenda in mind. Although this type of statement is not a complete fabrication, it is misleading.
 
 One type of statement often made on cherry-picked data is a trendline statement.
 By carefully selecting the start and the end point of the trendline, it is often possible to show misleading `trends' that are not representative of the real situation. These statements are prevalent in various fields, and particularly in  political contexts.  Climate change, in particular, has seen a great deal of this because intra-day and seasonal variations in temperature are so much greater than the global warming trend. 
 Let us look at several examples.
 
\begin{example}[Northern Hemisphere's Temperature]\label{example1}
In~\cite{temperature}, John Mason exhibits an example of using cherry-picked data points in the monthly temperature trendline dataset to distort the reality of climate change. By selecting a shorter time frame and cherry-picking specific locations, one can come out with the fantasy-like statement that: {\em The northern hemisphere summers are colder than winters.} For example, a cherry-picked summer day of \textit{Ann Arbor (MI, USA) on Aug. 18} had an average temperature of 58$\degree F$, which is 8 degrees lower than its average temperature on \textit{Mar. 15} (a winter day). In fact, both of the seasonal aggregation results and the validation~\cite{asudeh2020detecting} indicate that such cherry-picked trendline statements are not a fair representation of the truth. 
\end{example}

Cherry-picked claims can aggravate public panic and lead to potentially dangerous outcomes in policy-making, as shown in the next example.

\begin{example}[President Trump's COVID-19 statement (\textit{from politifact.com)}] On May 24, as the number of deaths of COVID-19 reached 100,000, President Donald Trump tweeted "\textit{Cases, numbers and deaths are going down all over the Country!}" to build an optimistic attitude towards the pandemic. If we look at a short time window, it was true that in some states, the pandemic was easing; so the statement was not totally false, rather it was cherry-picked.
The opposite conclusion would have been reached considering most other time periods, and particularly, longer periods.
\end{example}
\vspace{-2mm}

In practice, common fact-checking methods include crowd-sourced or expert-based manual fact-checking (i.e., PolitiFact~\cite{politifact}, FactCheck~\cite{factcheck}), and automatic fact-checking relying on techniques like information retrieval~\cite{doan2012principles}, natural language processing~\cite{li2007danalix}, and graph theory~\cite{cohen2011computational}. See ~\cite{zhou2018fake} for a comprehensive review. In the context of data management, ~\cite{wu2017computational} use query perturbation to evaluate the sensibility of the statements. \cite{asudeh2020detecting} focus on the trendline statements, and proposes efficient exact and approximate algorithms works for both unconstrained and constrained trendline statements. 

In this demonstration, we present MithraDetective, a system for detection of cherry-picked trendlines. MithraDetective adopts the definitions and algorithms developed in ~\cite{asudeh2020detecting} for statement validation and alternative statement discovery. We provide users an interactive UI, where they can set up the statement parameters to validate how cherry-picked a statement is and find alternatives. The pre-loaded datasets cover various controversial topics, including COVID-19, employment, and climate change. 

In the rest of this demo description, we first provide our system details, including what the system does, the system implementation, user interface, and back-end algorithm in Section~\ref{sec:system}. In Section~\ref{sec:plan}, we overview our demonstration plan.

\section{System Details}\label{sec:system}

\begin{figure*}[t]
\centering
\includegraphics[scale=0.4]{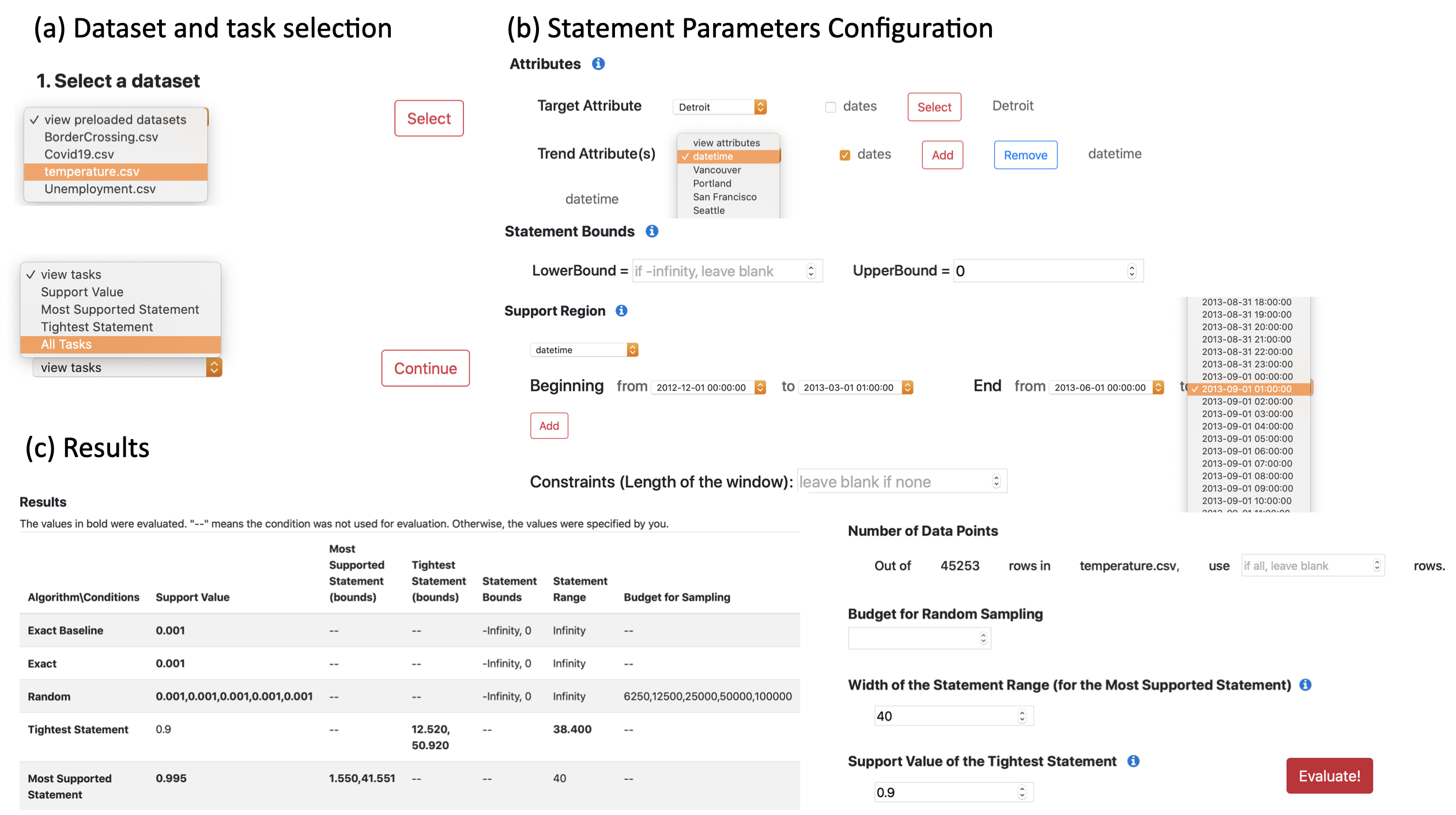}
\caption{MithraDetective: User Interface}\label{figure}
\end{figure*}

MithraDetective is a cherry-picked trendline statement fact-checking web application. Our focus is restricted to \textit{trendline statements} based on selected endpoints.  We perform both \textit{statement validation} and \textit{alternative discovery}.

\noindent
\textbf{Trendline Statement.} A trendline statement describes the relationship of a pair of trend points $b$ (beginning) and $e$ (end) and their target values, $y(b)$ and $y(e)$ respectively. A trendline statement provides a numerical upper bound and lower bound where $y(e) - y(b)$ falls within. For instance, in Example~\ref{example1}, the beginning point of the trendline is \textit{<Aug. 18, Ann Arbor>} and the end point is \textit{<March 15, Ann Arbor>}, the bound condition of the statement is $(0,\inf)$, which means that $y(e)- y(b) = 66 - 58 > 0$.

\noindent
\textbf{Statement Validation. }Given the region of all possible beginning points $R(b)$ and end points $R(e)$, the support of a statement is the proportion of $(b, e)$ pairs whose $y(e)- y(b)$ falls within the bounds of the statement.

\noindent
\textbf{Alternative Statement Discovery.} Given a threshold for the statement support score, for a statement whose score is lower than the given threshold, find an alternative statement that has (1) the highest support score and (2) the tightest statement boundary (see Section~\ref{sec:frontend} for details).

MithraDetective is built using the Flask framework (v1.1.1). The front-end is implemented using AngularJS (v1.8.0), Bootstrap (v4.0.0), HTML, and CSS. The back-end is written in Python 3.7.4. We describe the front-end user interface in Section~\ref{sec:frontend} and demonstrate the backend validation and alternative discovery algorithms in Section~\ref{sec:backend}.






\subsection{User Interface}\label{sec:frontend}
MithraDetective’s user interface has two input sections and one output section. 

\noindent
\textbf{Input: Dataset and Task Selection.} As shown in Figure 1(a), in the first input section, the user first selects a preloaded dataset related to the trendline of their interest. Then, they select one of the three tasks, or all, to perform: 
\begin{itemize}
  \item Support value: evaluates if a statement is cherry-picked by returning a value between 0 and 1. The higher the support value, the less cherry-picked the trendline is.
  \item Most supported statement (MSS): The user specifies a bound condition, which is the range of the upper and lower bounds. Given the condition, MSS returns the upper and lower bounds of a trendline with the highest support value. If your statement is cherry-picked, the most supported statement can be a better alternative.
  \item Tightest statement: returns the tightest bounds of a trendline statement with at least the given support value threshold.
\end{itemize}

\noindent
\textbf{Input: Statement Parameters Configuration.} As in Figure 1(b), in the second input section, the user specifies seven conditions for the evaluation of the trendline. (1) The user specifies the target and trend attributes, which define the trend points and their target values of the trendline. They should specify if an attribute represents dates. (2) Statement Bounds: the upper and lower bounds of a statement, within which $y(e)- y(b)$ would fall if the trendline supports the statement. (3) Support Region: A pair of disjoint regions, $R(b)$ and $R(e)$, to which every $b$ and $e$ belong respectively. Trendlines created from $b$ and $e$ are considered when computing the support value of the trendline statement of interest. The user should specify Constraints (Length of the window) if their trendline statement is constrained. (4) Budget for Random Sampling: The system improves the efficiency of the evaluation by using the random algorithm. The user specifies the number of points to sample from the Support Region. (5) Width of the Statement Range (for the MSS task): The user specifies a number greater than 0 for the range of the MSS. (6) Support Value of the Tightest Statement: The user specifies a number within $(0, 1)$. 

\noindent
\textbf{Output: Evaluation Results.} The output section, as shown in Figure 1(c),  displays the results of the evaluation as a table. In the table, the values in bold were the outputs of the algorithms, and those not in bold were conditions specified by the user as in Figure 1(b). The conditions that were not needed to complete the task are marked with "- -". 

If the user has specified the task as support value or all, the user sees the support values calculated using three algorithms: exact baseline, exact, and random. In Figure 1(c), the statement “Detroit summers are colder than winters” has been evaluated to have a support value of 0.001 for the exact baseline and exact algorithms. This means that the statement is barely supported by the points in the support region, and it is very likely cherry-picked. In addition, the statement has a support value of 0.001 for the random algorithm. If the user does not specify a budget, the system uses five large budgets shown in Figure 1(c), and calculates the support respectively. 

If the task is most supported statement or all, the user sees the upper and lower bounds of the statement range as well as the highest support value achieved. Using our Detroit example, we can state that the least cherry-picked statement with the statement range 40 has bounds (1.550, 41.551) and support value 0.995. In other words, “Detroit winters are colder than summers by 1.550 to 41.551 degrees” has the highest support value with 0.995, satisfying the condition that the statement range equals 40.

If the task is tightest statement or all, the user sees as the output the upper and lower bounds of the statement range which results in the specified support value. Again using our Detroit example, we can state that the statement that has the tightest statement range with support value 0.9 has bounds (12.520, 50.920) and range 38.400. "Detroit winters are colder than summers by 12.520 to 50.920 degrees” has at least 0.9 as its support value. The statement has the tightest range of 38.400, out of all statements with at least 0.9 support.


\subsection{Statement Validation and Discovery}\label{sec:backend}
For evaluating the support value, the system uses three algorithms. The \textit{exact baseline algorithm} uses the brute force algorithm. It first counts the total number of all possible pairs of the points in the beginning and end support regions. After counting the pairs that fulfill the support bounds, it divides the counts by the total number. To improve efficiency, the \textit{exact algorithm} uses binary search. If the dataset contains a very large number of records, the exact algorithm becomes inefficient. To tackle this problem, we use the \textit{random algorithm}, which uses the point sampling method. 
For evaluating the \textit{most supported statement} and \textit{tightest statement}, we also use efficient algorithms instead of brute force. The efficiency is improved by sorting the list of every $y(e)-y(b)$. 


\section{Demonstration Plan}\label{sec:plan}
MithraDetective is part of the Mithra system\footnote{\url{https://www.cs.uic.edu/~indexlab/projects.htm}}. We will demontrate it with three real-world datasets. They are focusing on controversial topics including COVID-19, employment, immigration policy, and climate change.
\begin{enumerate}
\item \textit{COVID-19\footnote{\url{https://covidtracking.com/}}}: \textit{COVID-19} dataset contains data related to COVID-19, a global pandemic which has affected the world significantly including the U.S.. There are 142 daily records for the U.S. from Apr. 1 to Aug. 20. The dataset includes 12 attributes, describing the numbers of tests with positive and negative results, persons in ICU, persons on ventilator, recovered, and deaths. 
\item \textit{Unemployment\footnote{\url{https://fred.stlouisfed.org/categories/32447}}}: \textit{Unemployment} dataset addresses the issue of unemployment situation in different gender, race, education-level, and age groups. There are 342 records of monthly unemployment rate for nine groups, published by the U.S. Bureau of Labor Statistics, from Jan. 1, 1992 to June 1, 2020.
\item \textit{BorderCrossing\footnote{\url{https://explore.dot.gov/views/BorderCrossingData/Monthly?:isGuestRedirectFromVizportal=y&:embed=y}}}: \textit{BorderCrossing} dataset contains data related to the number of border crossings in the U.S.-Canada and U.S.-Mexico borders. There are five attributes describing the method of crossing, port name, and state. The data is published by the Bureau of Transportation Statistics. 
\item \textit{Weather Dataset (WD)\footnote{\url{https://www.kaggle.com/selfishgene/historical-hourly-weather-data/home}}}: \textit{WD} contains 45,253 records of hourly temperatures for 35 cities in the United States, Canada, and Israel. For each city, the dataset contains temperatures in Kelvin from October 1, 2012 to November 30, 2017. We will demonstrate our system using Example~\ref{example1}.
\end{enumerate}

The user selects a dataset from the preloaded datasets. We use the \textit{Weather Dataset (WD)} to demonstrate how a user could interact with MithraDetective\footnote{A video for MithraDetective can be found at: \url{https://bit.ly/2EHOIOO}}.
\begin{enumerate}
    \item For the first section, select "\textit{temperature.csv}" from the drop-down menu, then click "Select." For the second section, select "All Tasks" from the drop-down menu, then click "Continue."
    \item Choose "\textit{Detroit}" for the target attribute and click "Select." Choose "datetime" for the trend attribute, check "dates," and click "Add." For the Statement Bounds, specify the upper bound as 0 and leave the lower bound blank. The lower bound is set to $-\infty$ by default. Next, choose "datetime" for the Support Region.  Specify the beginning from \textit{2012-12-01 00:00:00} to \textit{2013-03-01 01:00:00} and the end from \textit{2013-06-01 00:00:00} to \textit{2013-09-01 01:00:00}, then click "Add." Leave blank Constraints (length of the window) in Support Region, Number of Data Points, and Budget for Random Sampling. For the Width of the Statement Range, enter $40$. Lastly, for the Support Value of the Tightest Statement, enter $0.9$. Click "Evaluate!."
    \item The results are displayed. Depending on the task completed, the user will see one or all of the support values, tightest statement, and most supported statement. 
    \item If the user wishes to evaluate a constrained trendline, they should enter some value in Constraints under Support Region. If they wish to use the first $n$ rows of the dataset for evaluation, they should enter the number in Number of Data Points. $n$ must be a positive number less than the total number of rows, which is shown on the page. If they wish to specify the number of sampled points used in the Support Random algorithm, they should enter a positive number in the Budget for Random Sampling. 

\end{enumerate}



\bibliographystyle{ACM-Reference-Format}
\bibliography{ref}


\begin{thebibliography}{9}


\ifx \showCODEN    \undefined \def \showCODEN     #1{\unskip}     \fi
\ifx \showDOI      \undefined \def \showDOI       #1{#1}\fi
\ifx \showISBNx    \undefined \def \showISBNx     #1{\unskip}     \fi
\ifx \showISBNxiii \undefined \def \showISBNxiii  #1{\unskip}     \fi
\ifx \showISSN     \undefined \def \showISSN      #1{\unskip}     \fi
\ifx \showLCCN     \undefined \def \showLCCN      #1{\unskip}     \fi
\ifx \shownote     \undefined \def \shownote      #1{#1}          \fi
\ifx \showarticletitle \undefined \def \showarticletitle #1{#1}   \fi
\ifx \showURL      \undefined \def \showURL       {\relax}        \fi
\providecommand\bibfield[2]{#2}
\providecommand\bibinfo[2]{#2}
\providecommand\natexlab[1]{#1}
\providecommand\showeprint[2][]{arXiv:#2}

\bibitem[\protect\citeauthoryear{??}{fac}{2003}]%
        {factcheck}
 \bibinfo{year}{2003}\natexlab{}.
\newblock \bibinfo{title}{FactCheck}.
\newblock
\newblock
\newblock
\shownote{\url{https://www.factcheck.org/}.}


\bibitem[\protect\citeauthoryear{??}{pol}{2007}]%
        {politifact}
 \bibinfo{year}{2007}\natexlab{}.
\newblock \bibinfo{title}{PolitiFact}.
\newblock
\newblock
\newblock
\shownote{\url{https://www.politifact.com/}.}


\bibitem[\protect\citeauthoryear{Asudeh, Jagadish, Wu, and Yu}{Asudeh
  et~al\mbox{.}}{2020}]%
        {asudeh2020detecting}
\bibfield{author}{\bibinfo{person}{Abolfazl Asudeh}, \bibinfo{person}{H.~V.
  Jagadish}, \bibinfo{person}{You Wu}, {and} \bibinfo{person}{Cong Yu}.}
  \bibinfo{year}{2020}\natexlab{}.
\newblock \showarticletitle{On detecting cherry-picked trendlines}.
\newblock \bibinfo{journal}{\emph{PVLDB}} \bibinfo{volume}{13},
  \bibinfo{number}{6} (\bibinfo{year}{2020}), \bibinfo{pages}{939--952}.
\newblock


\bibitem[\protect\citeauthoryear{Cohen, Hamilton, and Turner}{Cohen
  et~al\mbox{.}}{2011}]%
        {cohen2011computational}
\bibfield{author}{\bibinfo{person}{Sarah Cohen}, \bibinfo{person}{James~T
  Hamilton}, {and} \bibinfo{person}{Fred Turner}.}
  \bibinfo{year}{2011}\natexlab{}.
\newblock \showarticletitle{Computational journalism}.
\newblock \bibinfo{journal}{\emph{Commun. ACM}} \bibinfo{volume}{54},
  \bibinfo{number}{10} (\bibinfo{year}{2011}), \bibinfo{pages}{66--71}.
\newblock


\bibitem[\protect\citeauthoryear{Doan, Halevy, and Ives}{Doan
  et~al\mbox{.}}{2012}]%
        {doan2012principles}
\bibfield{author}{\bibinfo{person}{AnHai Doan}, \bibinfo{person}{Alon Halevy},
  {and} \bibinfo{person}{Zachary Ives}.} \bibinfo{year}{2012}\natexlab{}.
\newblock \bibinfo{booktitle}{\emph{Principles of data integration}}.
\newblock \bibinfo{publisher}{Elsevier}.
\newblock


\bibitem[\protect\citeauthoryear{Li, Chaudhuri, Yang, Singh, and Jagadish}{Li
  et~al\mbox{.}}{2007}]%
        {li2007danalix}
\bibfield{author}{\bibinfo{person}{Yunyao Li}, \bibinfo{person}{Ishan
  Chaudhuri}, \bibinfo{person}{Huahai Yang}, \bibinfo{person}{Satinder Singh},
  {and} \bibinfo{person}{H.~V. Jagadish}.} \bibinfo{year}{2007}\natexlab{}.
\newblock \showarticletitle{Danalix: a domain-adaptive natural language
  interface for querying xml}. In \bibinfo{booktitle}{\emph{SIGMOD}}.
  \bibinfo{pages}{1165--1168}.
\newblock


\bibitem[\protect\citeauthoryear{Mason}{Mason}{2013}]%
        {temperature}
\bibfield{author}{\bibinfo{person}{John Mason}.}
  \bibinfo{year}{2013}\natexlab{}.
\newblock \bibinfo{title}{How to use short timeframes to distort reality: a
  guide to cherrypicking.}
\newblock
\newblock
\newblock
\shownote{\url{https://www.skepticalscience.com/cherrypicking-guide.html}.}


\bibitem[\protect\citeauthoryear{Wu, Agarwal, Li, Yang, and Yu}{Wu
  et~al\mbox{.}}{2017}]%
        {wu2017computational}
\bibfield{author}{\bibinfo{person}{You Wu}, \bibinfo{person}{Pankaj~K Agarwal},
  \bibinfo{person}{Chengkai Li}, \bibinfo{person}{Jun Yang}, {and}
  \bibinfo{person}{Cong Yu}.} \bibinfo{year}{2017}\natexlab{}.
\newblock \showarticletitle{Computational fact checking through query
  perturbations}.
\newblock \bibinfo{journal}{\emph{TODS}} \bibinfo{volume}{42},
  \bibinfo{number}{1} (\bibinfo{year}{2017}), \bibinfo{pages}{1--41}.
\newblock


\bibitem[\protect\citeauthoryear{Zhou and Zafarani}{Zhou and Zafarani}{2018}]%
        {zhou2018fake}
\bibfield{author}{\bibinfo{person}{Xinyi Zhou} {and} \bibinfo{person}{Reza
  Zafarani}.} \bibinfo{year}{2018}\natexlab{}.
\newblock \showarticletitle{Fake news: A survey of research, detection methods,
  and opportunities}.
\newblock \bibinfo{journal}{\emph{arXiv preprint arXiv:1812.00315}}
  (\bibinfo{year}{2018}).
\newblock


\end{thebibliography}
\end{document}